\documentclass[twocolumn]{aastex63}
\usepackage{verbatim}
\usepackage{comment} 
\def\tl{t_{\rm look}}

\makeatletter
\def\@hex@@Hex#1%
 {\if a#1A\else \if b#1B\else \if c#1C\else \if d#1D\else
  \if e#1E\else \if f#1F\else #1\fi\fi\fi\fi\fi\fi \@hex@Hex}
\makeatother
\definecolor{afcolor}{HTML}{b3443c}

\newcommand\quotes[1]{``{#1}"}
\newcommand\code[1]{\textsc{\MakeLowercase{#1}}}

\received{\today}
\shorttitle{The star formation history of GS-z14-0}
\shortauthors{A. Ferrara}
\graphicspath{{./}{figures/}}

\begin{document}



\def\be{\begin{equation}}
\def\ee{\end{equation}}
\def\gsim{\lower.5ex\hbox{\gtsima}} 
\def\lsim{\lower.5ex\hbox{\ltsima}} 
\def\gtsima{$\; \buildrel > \over \sim \;$} 
\def\ltsima{$\; \buildrel < \over \sim \;$} \def\gsim{\lower.5ex\hbox{\gtsima}} 
\def\lsim{\lower.5ex\hbox{\ltsima}} 
\def\simgt{\lower.5ex\hbox{\gtsima}} 
\def\simlt{\lower.5ex\hbox{\ltsima}}

\def\msun{{\rm M}_{\odot}}
\def\lsun{{\rm L}_{\odot}}
\def\dsun{{\cal D}_{\odot}}
\def\fsun{\xi_{\odot}}
\def\zsun{{\rm Z}_{\odot}}
\def\msunyr{\msun {\rm yr}^{-1}}
\def\gdens{\msun\,{\rm kpc}^{-2}}
\def\sfrdens{\msun\,{\rm yr}^{-1}\,{\rm kpc}^{-2}}

\def\mum{\mu {\rm m}}
\def\cc{\rm cm^{-3}}
\def\uflux{{\rm erg}\,{\rm s}^{-1} {\rm cm}^{-2} }

\def\fdust{\xi_{d}}
\def\fesc{f_{\rm esc}\,}
\def\td{\tau_{sd}}
\def\Sg{$\Sigma_{g}$}
\def\S*{$\Sigma_{\rm SFR}$}
\def\Ssfr{\Sigma_{\rm SFR}}
\def\Sgas{\Sigma_{\rm g}}
\def\Sstar{\Sigma_{\rm *}}
\def\Sesc{\Sigma_{\rm esc}}
\def\Srad{\Sigma_{\rm rad}}

\def\Dsolar{${\cal D}/\dsun$}
\def\Zsolar{$Z/\zsun$}
\def\DDsolar{\left( {{\cal D}\over \dsun} \right)}
\def\ZZsolar{\left( {Z \over \zsun} \right)}
\def\kms{{\rm km\,s}^{-1}\,}
\def\skms{$\sigma_{\rm kms}\,$}

\def\Scii{$\Sigma_{\rm [CII]}$}
\def\Sciimax{$\Sigma_{\rm [CII]}^{\rm max}$}
\def\CII{\hbox{[C~$\scriptstyle\rm II $]~}}
\def\CIII{\hbox{C~$\scriptstyle\rm III $]~}}
\def\OII{\hbox{[O~$\scriptstyle\rm II $]~}}
\def\OIII{\hbox{[O~$\scriptstyle\rm III $]~}}
\def\HH{\hbox{H$_2$}~} 
\def\HI{\hbox{H~$\scriptstyle\rm I\ $}} 
\def\HII{\hbox{H~$\scriptstyle\rm II\ $}} 
\def\CIion{\hbox{C~$\scriptstyle\rm I $~}}
\def\CIIion{\hbox{C~$\scriptstyle\rm II $~}}
\def\CIIIion{\hbox{C~$\scriptstyle\rm III $~}}
\def\CIVion{\hbox{C~$\scriptstyle\rm IV $~}}
\def\nhh{n_{\rm H2}}
\def\nhi{n_{\rm HI}}
\def\nhii{n_{\rm HII}}
\def\fhh{x_{\rm H2}}
\def\fhi{x_{\rm HI}}
\def\fhii{x_{\rm HII}}
\def\fd{f^*_{\rm diss}} 
\def\ks{\kappa_{\rm s}}

\def\cyan{\color{cyan}}
\definecolor{apcolor}{HTML}{b3003b}
\definecolor{afcolor}{HTML}{800080}
\definecolor{lvcolor}{HTML}{DF7401}
\definecolor{mdcolor}{HTML}{01abdf} 
\definecolor{cbcolor}{HTML}{ff0000}
\definecolor{sccolor}{HTML}{cc5500} 
\definecolor{sgcolor}{HTML}{00cc7a}

\title{The eventful life of GS-z14-0, the most distant galaxy at redshift $z=14.32$}\correspondingauthor{Andrea Ferrara}
\email{andrea.ferrara@sns.it}
\author[0000-0002-9400-7312]{Andrea Ferrara}
\affil{Scuola Normale Superiore,  Piazza dei Cavalieri 7, 50126 Pisa, Italy}

\begin{abstract}
We developed a model for the star formation history (SFH) of super-early galaxies and applied it to GS-z14-0, the most distant galaxy known, located at $z=14.32$ (294 million years after the Big Bang). The SFH, starting at $z=26.7$, is complex. Initially ($z>18$), the galaxy experiences feedback-regulated phases that are bursty, relatively faint (reaching $M_{\rm UV}=-18.4$), and unattenuated. When dust shielding allows for a smooth star formation rate (SFR), the galaxy quickly becomes heavily obscured. During this obscured phase, which lasts for approximately 20\% of the total star-forming time, 70\% of the observed stars are formed. Super-early galaxies in this phase should be detectable by ALMA. 
Twenty-six million years before observation, as the galaxy becomes super-Eddington, a powerful radiation-driven outflow clears most of the dust and significantly reduces the SFR by a factor of seven, from $100 \to 15\ \msunyr$. The galaxy transitions into a \quotes{blue monster} dominating the bright end of the UV luminosity function. When the outflow ceases due to decreased dust opacity, the galaxy relaxes into a \textit{post-starburst} phase, in which it is currently observed.
Our model accurately reproduces all the observed and inferred properties of the galaxy. The analysis of this extreme system opens exciting opportunities for studying the beginnings of the luminous Universe.
\end{abstract}
\keywords{galaxies: high-redshift, galaxies: evolution, galaxies: formation}

\section{Introduction} \label{sec:intro}
The \textit{James Webb Space Telescope} (JWST), by breaking the redshift $z=10$ barrier, is finally pushing the frontier of the known universe deep into the Cosmic Dawn, the epoch when the first stars and galaxies formed. One of the major puzzles raised by these data \citep{Bouwens22b, Harikane23, Harikane23c, McLeod23, Finkelstein23, Adams23, Willott24, Perez23, Robertson23} concerns the overabundance of very bright, blue (and sometimes massive for their epoch) galaxies boosting the bright end of the UV luminosity function (LF).

The overabundance of such \quotes{blue monsters} has been explained, among other suggestions\footnote{These include (a) star  formation variability \citep{Mason23,Mirocha23,Pallottini23}, (b) reduced feedback resulting in a higher star formation efficiency \citep{Dekel23,Li23}, and (c) a top-heavy IMF \citep{Inayoshi22}, although see \citet{Cueto23}.}, as a result of negligible dust attenuation in super-early galaxies \citep{Ferrara23a, Fiore23, Ziparo23, Ferrara24}. Clear signs of a decreasing UV slope with redshift, $\beta$, have been reported \citep[e.g.][]{Cullen23, Austin24, Stiavelli23}. Indeed, this simple hypothesis produces extremely good fits to the observed UV LFs at $z>10$ \citep{Ferrara24, Donnan24}. 

However, this interpretation is at odds with the expectation that significant amount of dust, and hence attenuation, should be present given the large stellar masses ($10^{8-9}\ \msun$) and compact sizes ($\approx 100$ pc, \citealt{Ono22, Sun23, Baldwin24, Morishita24, Langeroodi24}) often observed in these systems. As a potential solution, it has been proposed \citep{Ferrara24} that the intense stellar luminosity, coupled with the dust opacity, might render the galaxy super-Eddington, and able to drive powerful outflows that disperse the dust in the circumgalactic medium, making the galaxy visible and UV-bright. 

To consolidate this scenario, it is necessary to study its implications. In particular, it should be able to explain not only the dust production and ejection, but also the star formation histories (SFH) and observed properties of blue monsters. 

Here, we examine the outstanding case of the newly discovered, spectroscopically confirmed galaxy GS-z14-0 at $z=14.32$ \citep{Carniani24} found by the \textit{JWST} JADES survey \citep{Eisenstein23}. The galaxy has also been detected in the wide 7.7 $\mu$m MIRI filter \citep{Helton24}. With its bright luminosity, $M_{\rm UV}=-20.81$, blue colors, $\beta=-2.2$, and small dust attenuation, $A_V=0.31$, GS-z14-0 located only 294 Myr after the Big Bang, represents a fascinating yet very challenging test case for the scenario described above.

We will see that super-early galaxies show complex SFH, and undergo bursty, obscured and \quotes{mini-quenching} star formation phases regulated by feedback processes, finally leading, at least in the case of GS-z14-0, to a blue monster. Thus, the analysis of this extreme galaxy, opens new and exciting perspectives for studying the beginnings of the luminous Universe\footnote{Throughout the paper, we assume a flat Universe with the following cosmological parameters: $\Omega_{\rm M} = 0.3075$, $\Omega_{\Lambda} = 1- \Omega_{\rm M}$, and $\Omega_{\rm b} = 0.0486$,  $h=0.6774$, $\sigma_8=0.826$, where $\Omega_{M}$, $\Omega_{\Lambda}$, and $\Omega_{b}$ are the total matter, vacuum, and baryon densities, in units of the critical density; $h$ is the Hubble constant in units of $100\,\kms$, and $\sigma_8$ is the late-time fluctuation amplitude parameter \citep{planck:2015}.}.   

\section{Basic equations}\label{sec:Equations}
To derive the feedback-regulated SFH of super-early galaxies we start by writing the 
equations\footnote{We assume the Instantaneous Recycling Approximation (IRA).} \citep[see, e.g.][]{Ferrara00} determining the evolution of dark matter halo mass, $M$, gas mass, $M_g$, stellar mass, $M_\star$, and metal
mass, $M_Z$ with cosmic time, $t$, in the galaxy:
\begin{eqnarray}
&\dot M(t)& = \dot M_a(M,t)\label{eq:M}\\
&\dot M_g(t)& = f_b \dot M_a -[(1-R)+\eta(t-t_D)] \dot M_\star(t)  \label{eq:Mg}\\
&\dot M_\star(t)& = \epsilon_\star \frac{M_g(t)}{t_{\rm ff}} f_{\rm LW}(t-t_D) (1-R) \label{eq:Ms}\\    
&\dot M_Z(t)& = \{[\xi_Z-Z(t)](1-R)-\eta(t-t_D)\} \dot M_\star(t).  \label{eq:MZ}
\end{eqnarray}
The halo growth is fed by cosmological accretion, whose rate (in $\msunyr$) is obtained from numerical simulations, e.g. \citet{Correa15}:
\begin{equation}
\dot M_a(M,z) = 102.2\, h [-\alpha -\beta(1+z)] E(z) M_{12} \label{eq:Macc}  
\end{equation}
where $(\alpha, \beta)$ are parameters that depend on $M(z)$, cosmology and the linear matter power spectrum, and provided\footnote{For reference, when averaged over $8 < \log (M_0/\msun) < 14$, $(\alpha, \beta)=(0.25, -0.75)$.} in Appendix C of \citet{Correa15}; finally, $E(z) = [\Omega_m(1+z)^3+\Omega_\Lambda)]^{1/2}$, and $M_{12}= M/10^{12} \msun$. We assume that no stars form before redshift $z = z_8$ at which the halo crosses the atomic-cooling limit mass, $M(z_8)\approx 10^8 \msun$ \citep{Salvadori09, Ferrara14}. The results are however insensitive to the precise choice of this value.

\subsection{Gas evolution}
Gas is (i) accreted at the same rate as in eq. \ref{eq:Macc} but scaled by the cosmological baryon fraction\footnote{We assume that accreted gas is metal-free} $f_b = \Omega_b/\Omega_m = 0.158$, (ii) consumed by star formation,  and (iii) lost via galactic outflows. These processes are described by the three r.h.s. terms in eq. \ref{eq:Mg}, respectively. The outflow rate is equal to $\eta \dot M_\star$, where the loading factor $\eta$ includes a delay time, $t_D$. In principle $t_D$ could be different for the different types of feedback \citep{Pallottini23}, or depend on galaxy properties such as size or stellar mass. For simplicity, we adopt a fixed value of $t_D$ which is left as a free parameter of the model. The derivation of $\eta$ for radiation-driven outflows is given in Sec. \ref{subsec:outflow}. We adopt a $1-100\, \msun$ Salpeter IMF, for which the return fraction from stars, assuming IRA, is $R=0.69$. 

\subsection{Star formation}
Eq. \ref{eq:Ms} describes the conversion of gas mass into stars. The process occurs with an \textit{instantaneous} efficiency, $\epsilon_\star$, per free-fall time, $t_{\rm ff} = \zeta H(z)^{-1}$, where $\zeta = 0.06$ and $H(z)^{-1}$ is the Hubble time.  We implicitly assume that the efficiency is regulated by the mechanical energy input by SNe for which we adopt the functional form in \citet{Ferrara23a}:
\be\label{eq:eps}
\epsilon_\star = \epsilon_0\, \frac{v_c^2}{v_c^2+ f_w v_s^2}  
\ee
where $v_c$ is the halo circular velocity, $f_w=0.12$ is the SN energy coupling efficiency with the gas, and $v_s = \sqrt{\nu E_0} = 975\, \kms$ is a characteristic velocity associated with the SN energy released per unit stellar mass formed; $E_0=10^{51}\, \rm erg$, $\nu^{-1} = 52.89\, \msun$, for the adopted IMF. We fix $\epsilon_0=0.07$, a value constrained by the fit to the luminosity functions at $z\simgt 10$ \citep{Ferrara23a, Ferrara24}. Star formation is also limited (via a suppression factor $0 < f_{\rm LW} < 1$)  by the availability of molecular hydrogen which is destroyed by photons in the Lyman-Werner (LW) energy range ($11.2-13.6$ eV) with a delay time $t_D$. Such radiative feedback will be described in Sec. \ref{subsec:LW}. 

\subsection{Metal enrichment}
Eq. \ref{eq:MZ} describes the evolution of the metal mass $M_Z$. Heavy elements are produced by stars, partly recycled into stars and ejected by outflows. The IMF-integrated yield of heavy elements produced by stars is $\xi_Z = y_Z \nu$, where $y_Z$ is the SN yield. We compute $y_Z$ from the results in \cite{Woosley07}, and find $y_Z=2.41\, M_\odot$, with very weak dependence on stellar metallicity. 

\subsection{Dust}
From the metal and gas mass we derive the gas metallicity, $Z=M_Z/M_g$ that we use to compute the (normalized) dust-to-gas ratio , ${\cal D} = (M_d/M_g){\cal D}_{\rm MW}^{-1}$, where $M_d$ is the dust mass, and   ${\cal D}_{\rm MW}=1/162$ is the Milky Way value.  To this aim, we follow \cite{Remy14} who inferred $\cal D$ for a large sample of local galaxies over a wide metallicity range. They find  ${\cal D} = (Z/Z_\odot)^\alpha$, with $\alpha = 1.62 \pm 0.34$;   

From the dust mass we derive the galaxy dust optical depth. Following \citet{Ferrara24}, we write 
\begin{equation}
\tau_{1500} = \frac{\kappa_{1500}}{4\pi r_e^2}  M_d,    
\label{eq:tau1500}
\end{equation}
where the dust mass absorption coefficient at 1500 \AA\, is $\kappa_{1500} = 1.26\times 10^5\ {\rm cm}^2 {\rm g}^{-1}$ for a Milky Way extinction curve.  To determine the stellar effective radius, $r_e$, we follow \citet[see also \citealt{Morishita24}]{Shibuya15}, and take $r_e(z) = 0.01\, r_{\rm vir}(z)$.  For reference, a $10^{12} M_\odot$ halo at $z=14.32$ has $r_e = 210$ pc.

\section{Feedback processes}\label{sec:Physics}
The evolution of the galaxy properties (gas, stellar, metal, dust mass) described by eqs. \ref{eq:Mg}$-$\ref{eq:MZ} is strongly regulated by feedback processes. We explicitly consider two feedback types, namely (i) Lyman-Werner radiative feedback, and (ii) radiation-driven outflows. 

SN feedback implicitly regulates star formation by controlling the efficiency (eq. \ref{eq:eps}). However, the SN contribution to outflow driving is subdominant, and not included in the model. The main reason is that at the high gas densities characterising super-early galaxies \citep{Isobe23}, SN kinetic energy is very rapidly radiated away by catastrophic cooling  of the post-shock gas \citep{Pizzati20}, as discussed  in \citet{Ferrara24}. 

Feedback cannot be instantaneous, for a variety of physical reasons related to stellar evolution, radiation field fluctuations, evolution of photo-dissociated and ionized regions, gas dynamics and accretion. Each of these processes acts on a specific timescale \citep{Pallottini23}.  Hence, it is important to include the feedback delay in the model, at least in a simplified manner. As we will see, delayed feedbacks introduce a physically rich SFH behavior.

\subsection{Lyman-Werner feedback}\label{subsec:LW}
Star formation is almost always associated with cold, molecular gas. However, the presence of molecules is not a necessary requirement \citep{Glover12}. Rather, the key requirement is that the gas can self-shield from UV radiation, a condition favored by the presence of dust and molecules. Thus, molecule formation and the formation of cold gas are both correlated with the column density of the cloud, and thus its shielding ability.

UV radiation in the LW band is absorbed by \HH molecules (and dust). A fraction $\fd\approx 0.15$ of the absorptions \citep{Krumholz08} leads to photo-dissociation of the molecule, via the 2-step Solomon effect \citep{Sternberg14}. 

To quantify the effects of LW feedback we introduce the shielding radius, $r_{\rm sh}$, i.e. the radial distance at which the LW flux produced by the stars located at the center of the halo can penetrate, and hence inhibit star formation. We use the results in  \citet{Ferrara19} to write
\begin{equation}
r_{\rm sh} = \frac{N_d}{n} \ln(1+ w \chi). 
\label{eq:rsh}
\end{equation}
In eq. \ref{eq:rsh} $N_d$ is the dust column density at which the dust optical depth to UV photons (at 1000 \AA) becomes equal to unity, 
\be\label{eq: Nd}
N_d = 1/\sigma_d = 5\times 10^{20}  {\cal D}^{-1} \,\rm cm^{-2}.
\ee
Assuming that a fraction $f_d = 0.1$ of  the total gas mass is located within $r_e$, the mean gas density is  $n = 3 f_d M_g/4 \pi m_p r_e^3$. Finally,    
\begin{equation}\label{eq:w_dust}
w = \frac{1}{1+0.9 \mathcal{D}^{1/2}}, \qquad \chi = \frac{\fd \sigma_d} {\mathcal R}\frac{F_{\rm LW}}{n}.
\end{equation}
The first quantity is a correction factor accounting for the probability that a LW photon is absorbed by dust grains (associated with H$_2$) rather than by H$_2$ molecules. 

The parameter $\chi$ is instead proportional to the LW photo-dissociation parameter $F_{\rm LW}/n$, where $F_{\rm LW}$ is the LW photon flux [$\rm cm^{-2} s^{-1}$].  The coefficient contains the  total \HH formation rate in the gas phase catalyzed by $H^-$,  and on grain surfaces:  ${\cal R} = (k_{\rm H^-}x_e + k_d)$.  The reaction rates depend on the gas temperature, $T$, and ionization fraction, $x_e$, taken to be $10^2$ K and $10^{-4}$, respectively; they are written\footnote{We use the compact notation $Y_x = Y/10^x$}  as     $k_{\rm H^-}=10^{-16} T_2\, \rm cm^3 s^{-1}$ \citep{bovino:2016},  and $k_d = 3\times 10^{-17} T_2^{1/2} {\cal D}\, \rm cm^3 s^{-1}$ \citep{Cazaux04}. We note that \HH formation on grains dominates when the dust-to-gas ratio is ${\cal D}\simgt 10^{-3}$. 

We compute  the LW flux at time $t$,  as $F_{\rm LW}(t) = \dot N_{\rm LW}(t)/4\pi r_S^2(t)$, i.e. at the edge of the Str\"omgren sphere (of radius $r_S$)  produced by the stars, beyond which the gas becomes neutral.  To determine $r_S$ we derive the intrinsic galaxy UV luminosity at 1500\AA, $L_{1500}(t)$, from the star formation rate (${\rm SFR} \equiv \dot M_\star$) via a conversion factor, ${\cal K}_{1500}$ $[{L_\odot}/M_\odot {\rm yr}^{-1}]$, whose value has been chosen so to match the one used by the ALMA REBELS survey \citep{Bouwens22a}: ${\cal K}_{1500} \equiv {L_{1500}}/{\rm SFR} = 0.587 \times 10^{10}$. We then convert the UV luminosity into an ionizing photon rate multiplying $L_{1500}$ by the ionization parameter, $\xi_{\rm ion}=10^{25.7}$ in erg$^{-1}$ Hz \citep{Bunker23}.   

The photon rate $\dot N_{\rm LW}(t)$  [s$^{-1}$] is computed by integrating over the SFH:
\be 
\dot N_{\rm LW}(t) =   \int_{t(z_8)}^t  {\rm SFR}(t') Q_{\rm LW}(t-t') dt',  
\label{eq:Ndot}
\ee 
where the LW band-averaged photon production rate, $Q_{\rm LW}$,  per solar mass of stars formed in a instantaneous burst at $Z=0.2 Z_\odot$ is taken from \texttt{Starburst99}\footnote{http://www.stsci.edu/science/starburst99/} \citep{Leitherer99}. 

It is also useful to introduce the specific LW luminosity (in erg s$^{-1}$ Hz$^{-1}$),  is computed by integrating the production rate over the SFH:
\be 
L_{\rm LW}(t) = \frac{h\langle \nu \rangle}{\Delta\nu} \dot N_{\rm LW}(t) 
\label{eq:LW}
\ee 
where $\langle \nu \rangle$ is the mean frequency of the LW energy band of width $\Delta\nu$. Finally, the LW intensity [erg s$^{-1}$ cm$^{-2}$  Hz$^{-1}$ sr$^{-1}$] at a distance  $r$ from the source and at time $t$ is given by:
\be
J_{\rm LW}(r, t) =\frac{L _{\rm LW} (t)}{16\pi^2 r^2}.
\label{eq:JLW}
\ee
This is often conveniently expressed as $J_{\rm LW} = 10^{-21} J_{21}\, \rm erg\, s^{-1} cm^{-2}  Hz^{-1} sr^{-1}$.  As a reference, we will usually give the value of $J_{21}$ at the virial radius of the galaxy, i.e. $J_{21}(r_{\rm vir},t)$. 

We finally define the feedback strength, $f_{\rm LW}$, as 
\be
f_{\rm LW}(t) =\max [0, (1-r_{\rm sh}/r_{\rm vir})].
\label{eq:fLW}
\ee
This choice is motivated by the fact that for an isothermal distribution of the gas in the halo gravitational potential, $M_g \propto r$. Hence, $f_{\rm LW}$ quantifies the fraction of the total gas mass in which star formation is inhibited by the effects of UV radiation. We also impose that LW feedback acts with a delay time, $t_D$, which represents a free parameter of our model that must be calibrated from data (see Sec. \ref{sec:Results}). As a consequence, the feedback strength is evaluated at a retarded time,  $f_{\rm LW}(t-t_D)$ in eq. \ref{eq:Ms}. 

\subsection{Radiation-driven outflows}\label{subsec:outflow}
In addition to suppressing star formation, radiation can drive outflows by transferring the momentum of the photons to the dust and gas (these two components are tightly coupled by viscous and Coulomb drag forces). In a dusty medium, the classical Eddington luminosity $L_{E}=4\pi G m_p c M_*/\sigma_T = 1.26\times 10^{38}(M_*/M_\odot)\, \rm erg\, s^{-1}$ is reduced by a boost factor $A = \sigma_d/\sigma_T$, the ratio of the dust to Thomson cross-section, with $A \simeq 100-1100$, depending on dust abundance and properties. Given these uncertainties, for simplicity we scale $A$ linearly with dust-to-gas ratio, i.e. $A({\cal D}) = 1000 ({\cal D}/{\cal D_{\rm MW}})$, deferring a more precise determination to future work (Nakazato et al. in prep).  

\citet[][see also \citealt{Fiore23}]{Ferrara24} have shown that the modified super-Eddington condition, $L_{\rm bol} = f_{\rm bol}L_{1500} > A^{-1} L_E$, where the bolometric correction $f_{\rm bol} = 2$, can be cast into one on the specific star formation rate, ${\rm sSFR=SFR}/M_\star$: 
\begin{equation}
{\rm sSFR} > {\rm sSFR}^\star \simeq 25 \left(\frac{100}{A}\right) {\rm Gyr^{-1}}.
\label{eq:ssfr_thresh}
\end{equation}
Eq. \ref{eq:ssfr_thresh} represents the necessary condition for a galaxy to develop a radiation-driven outflow as soon as the galaxy sSFR crosses the sSFR$^\star$ threshold. 

We define the outflow rate as $W(t) = \eta \dot M_\star(t)$, with the mass loading factor $\eta =0$ if the galaxy is in the sub-Eddington regime, $\rm sSFR \le sSFR^\star$; $\eta$ is, in principle, a time-dependent function of several quantities (see below). The mass loading factor is included  into eq. \ref{eq:Mg} and \ref{eq:MZ}, where it is evaluated at a retarded time $t-t_D$. As a precise determination of $\eta$ is difficult both observationally and from numerical simulations \citep[][]{Muratov15, Pallottini22}, we choose to leave it as a free parameter of the model. Nevertheless, it is instructive to obtain at least a crude estimate of $\eta$ from physical arguments. 

The momentum injection rate due to stellar radiation is 
\be\label{eq:W}
W(t) v(t) =  f(\tau) \frac{L_{\rm bol}(t)}{c}, 
\ee
where $v$ is the outflow shell velocity, and 
\be
f(\tau) = (1-e^{-\tau_{1500}}) + \tau_{\rm IR}
\label{eq:ftau}
\ee
is a function of the UV (1500 \AA) and IR optical depths \citep{thompson2015, Ishibashi2018, Costa18}. The first term in eq. \ref{eq:ftau} accounts for single-scattering momentum transfer; the second one quantifies the momentum exchange between trapped IR radiation and the outflow. By substituting the expression for $L_{\rm bol} =f_{\rm bol} {\cal K}_{1500} {\rm SFR}$, and recalling the definition of $\eta$ above, we find
\be\label{eq:eta}
\eta \equiv \frac{W}{\rm SFR} =  236 f(\tau) \left(\frac{v}{\kms} \right)^{-1}.
\ee
Hence, large values of $\eta$ are achievable if $f(\tau) \gg 1$ and/or for low velocities. These conditions are likely to be met at the onset of the outflow, when the galaxy just became super-Eddington. 

Additionally, it is noteworthy that very large values, up to $\eta \approx 1000$, are also observed for classical, SN-driven outflows in low-mass galaxies \citep{Muratov15}. However, it is advisable not to extend this comparison further, considering the significant differences in the underlying physics between these two cases. A precise determination of $\eta$ is beyond the scope of the present study and necessitates radiation-hydrodynamics simulations, which we defer to future work.

%
%
\begin{figure*}
\centering\includegraphics[width = 1.0 \textwidth]{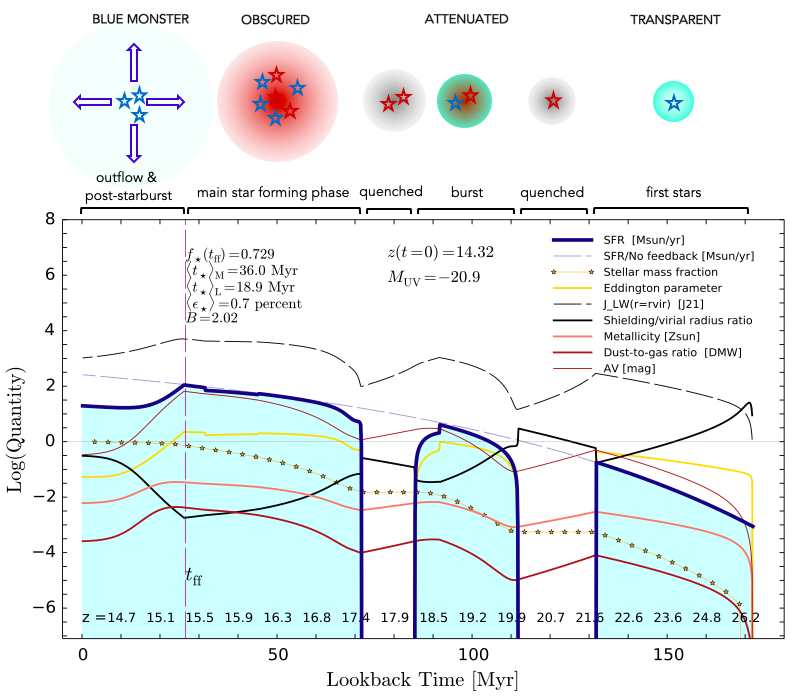}
\caption{Evolutionary tracks of key physical quantities of GS-z14-0 as a function of the lookback time from $z=14.32$ corresponding to the fiducial values of the free parameters: $M=8\times 10^{11}\, \msun$, $t_D = 40\, \rm Myr$, $\eta = 158$. As indicated in the legend, we show the SFH (blue thick line/shaded area), SFH with no feedback (gray dashed), the stellar mass fraction with respect to its final value (starred yellow), the Eddington parameter $\equiv\rm SFR/SFR^\star$  (thick yellow), the LW intensity at $r_{\rm vir}$ (dashed black), the shielding/virial ratio, $r_{\rm sh}/r_{\rm vir}$ (thick black), the metallicity (orange), dust-to-gas ratio, and $A_V$. The units for each quantity are reported in the legend. Also shown are the derived values for the fraction of stars formed before a free-fall time at $z=14.32$, $f_\star(t_{\rm ff})$, the mass- (luminosity-) weighted mean ages of the stellar population, $\langle t \rangle_M$ ($\langle t \rangle_L$), the gas-to-stars conversion factor, $\langle \epsilon_\star \rangle$, and the Balmer break strength, $B$. The top sketch schematically represents the evolutionary phases discussed in Sec. \ref{sec:Results}.} 
\label{Fig:SFH}
\end{figure*}
%
%

%
\begin{table*}
\centering
\caption{Measured and model predicted properties of the galaxy GS-z14-0 at $z =14.32$ }
\begin{tabular}{lccccccccc}
\hline\hline
 &$M_{\rm UV}$ & $\beta$ &$r_e$ & $\log (M_\star/M_\odot)$ & $\rm SFR_{10}$ & $A_V$ & $\log(Z/Z_\odot)$ & $\langle t \rangle_M$ & $\langle t \rangle_L$ \\
\hline 
 &      mag   & & pc    &                          &  $\msunyr$     &   mag &                   &   Myr  & Myr               \\
\hline
\textit{Measured} & $-20.81^{+0.16}_{-0.16}$ & $-2.2^{+0.07}_{-0.07}$ & $260^{+20}_{-20}$& $8.7^{+0.7}_{-0.2}$ & $22^{+6}_{-6}$           &   $0.31^{+0.14}_{-0.07}$ & $-1.5^{+0.7}_{-0.4}$  &  \textemdash & \textemdash \\
\hline
\textit{Predicted} &$-20.86$ & $-2.25$ &$196.2$  & $8.95$ & $19.6$           &   $0.34$ & $-2.2$  &  36.0 & 18.9 \\
\hline
\label{tab:properties}
\end{tabular}
\tablecomments{The \textit{Measured} row contains the observed or inferred properties of GS-z14-0 \citep{Carniani24}; the \textit{Predicted} row give the analogous results form our model.  Columns (2)-(10) give the galaxy UV magnitude and spectral slope, effective radius, stellar mass, SFR averaged over the last 10 Myr, visual extinction, gas metallicity, mass- and luminosity-averaged age of the stellar population, respectively.} 
\end{table*}

\section{Results}\label{sec:Results}
To avoid the awkwardness of general terms, and to elucidate some unique features of the SFH of super-early systems, we will use the recently discovered  $z=14.32$ (cosmic age 294 Myr) galaxy GS-z14-0 as a guiding example. However, the results can be used to understand the earliest galaxy population at large. 

With an observed UV luminosity $M_{\rm UV} = -20.8$, and a stellar mass $\log (M_\star/\msun) \simeq  {8.7}$ \citep{Carniani24}, GS-z14-0 has a star formation rate (averaged over the last 10 Myr) $\rm SFR \approx  22\, \msunyr$. We use our model to highlight the key evolutionary features of this dazzling system. However, the results can be used to understand the earliest galaxy population at large.  

Fig. \ref{Fig:SFH} shows the SFH (blue line/shaded area) of GS-z14-0 as a function of the lookback time, $\tl$, from $z=14.32$. This track corresponds to the fiducial values of the free parameters: $M=8\times 10^{11}\, \msun$, $t_D = 40\, \rm Myr$, and $\eta = 158$. For comparison, we also show the SFH when both the LW and the outflow are turned-off (gray, dashed). We identify six evolutionary phases (a)-(f), which are dominated by different physical processes. They are discussed in the following.

\paragraph{(a) The onset of star formation}
The galaxy started to form stars as its halo reached a mass of $10^8 \msun$, thus enabling Ly$\alpha$ cooling of the gas; this happened at $\tl=170\, \rm Myr$ or $z_8=26.7$. Since then, in a time $\approx t_D$, the SFR increased from $10^{-3}$ to $0.3\, \msunyr$. Due to the parallel increase of the LW radiation field, reaching an intensity $J_{21} \approx 300$, star formation was totally quenched in the entire galaxy by the inability of gas to self-shield. The little amount of dust formed by this first SF episode, corresponding to ${\cal D} = 10^{-4}$, was in fact insufficient to provide the minimum opacity to prevent the quenching. The amount of stars produced by this first burst is only 0.05\% of the stars observed at $z=14.32$.

\paragraph{(b) The earliest quenched phase}
At $\tl = 130$ Myr the galaxy stop forming stars for about 20 Myr. During this period, cosmological accretion continued. The infall of virtually pristine, dust-free
gas diluted the metals and the dust, thus decreasing $Z$ ($\cal D$) by a factor $3$ ($6$). 

At the same time, the UV/LW radiation field became less intense as a result of the passive ageing of the stellar population. At $\tl = 110$ Myr, $J_{21}$ was weak enough that $r_{\rm sh} < r_{\rm vir}$, thus allowing for a re-ignition of star formation. At this stage the metallicity is $\approx 10^{-3}\, Z_\odot$, and the galaxy has always been optical thin ($A_V\ll 1$) throughout the previous evolution.

\paragraph{(c) From bursty to continuous star formation}
Due to the large reservoir ($M_g \approx 2\times 10^9\, M_\odot$) of gas accumulated during the quenched phase, the SFR starts again vigorously, and in about $20$ Myr it reaches the level expected in the absence of feedback ($4\, \msunyr$). We note that the star formation efficiency is nevertheless small, $\epsilon_\star = 3.5$\%. As a result, the metal and dust mass, along with $J_{21}$, increases again. However, their level is not yet sufficient to fully self-shield the galaxy as the retarded effect of the dust dilution due to accretion in phase (b) stops the SFR again. 

This second stop, however, is relatively short and lasts only for 15 Myr. After this second burst the stellar mass is $1$\% of its final value. Quite importantly, though, dust produced during this event has for the first time significantly obscured the galaxy ($A_V = 1-3$) before being slightly diluted again  by accretion. 

At the end of the second quenched period, when the halo mass is $M\approx 8\times 10^{10}\, \msun$ and the metallicity is $\approx 0.01\, Z_\odot$, the bursty phase ends, and the galaxy relaxes onto a continuous star formation mode. Hence, although remarkable, the initial bursty phases induced by the LW feedback leave little imprints on the final properties of the galaxy, as we will discuss below. However, galaxies in these initial ($z \simgt 18$) bursty phases can be detected by \textit{JWST}. In fact, the brightest magnitudes of the galaxy during the first and second burst are $M_{\rm UV}=-16.1$ (at $z=21.8$) and $M_{\rm UV}=-18.4$ (at $z=20.3$), thus opening the exciting perspective to observe the very first phases of galaxy formation, and even the first Pop III stars.  

The above finding in not necessarily in tension with the reported non-detection of galaxies at $z=16-20$ \citep{Robertson23}, as the predicted magnitudes are below the detection limit of their survey. In the stacked images they reach as deep 31.4 AB mag, which at $z=20.3$, where the brightest bursty phase of a galaxy like GS-z14-0 occurs, would correspond to a magnitude of $M_{\rm UV}=-20.5$, which is shallower than the predicted $M_{\rm UV}=-18.4$. Deeper surveys will be able to test this prediction.

\paragraph{(d) The main star-forming period}
As the galaxy grows in mass and size, from $\tl =70$ Myr the SFR approaches $17\, \msunyr$ and steadily increases up to a dazzling $100\, \msunyr$. The LW feedback is virtually de-activated by the ability of the galaxy to self-shield via dust opacity, and the galaxy basically follows the no-feedback track. The gas density is extremely high,  $n \simeq 3 \times 10^4\, \cc$, thus ensuring that SN kinetic energy is catastrophically radiated away. 

This phase marks the main star forming period of the galaxy during which about 70\% of all its stars are formed. We note that about one third  ($27$\%) of the final stellar mass is assembled in the last free-fall time, $t_{\rm ff}=26.3$ Myr, before observation. A key point is that \textit{this stellar build-up phase is completely obscured}, as the large dust mass ($M_d \simeq 1.7\times 10^6\, \msun$) and concentration ($r_e = 150$ pc) at $\tl = t_{\rm ff}$ produces a very large ($A_V \simgt 50$) attenuation of the galaxy luminosity, rendering it essentially unobservable to \textit{JWST} (but possibly not to ALMA).  

%
%
\begin{figure*}
\centering\includegraphics[width = 1.0 \textwidth]{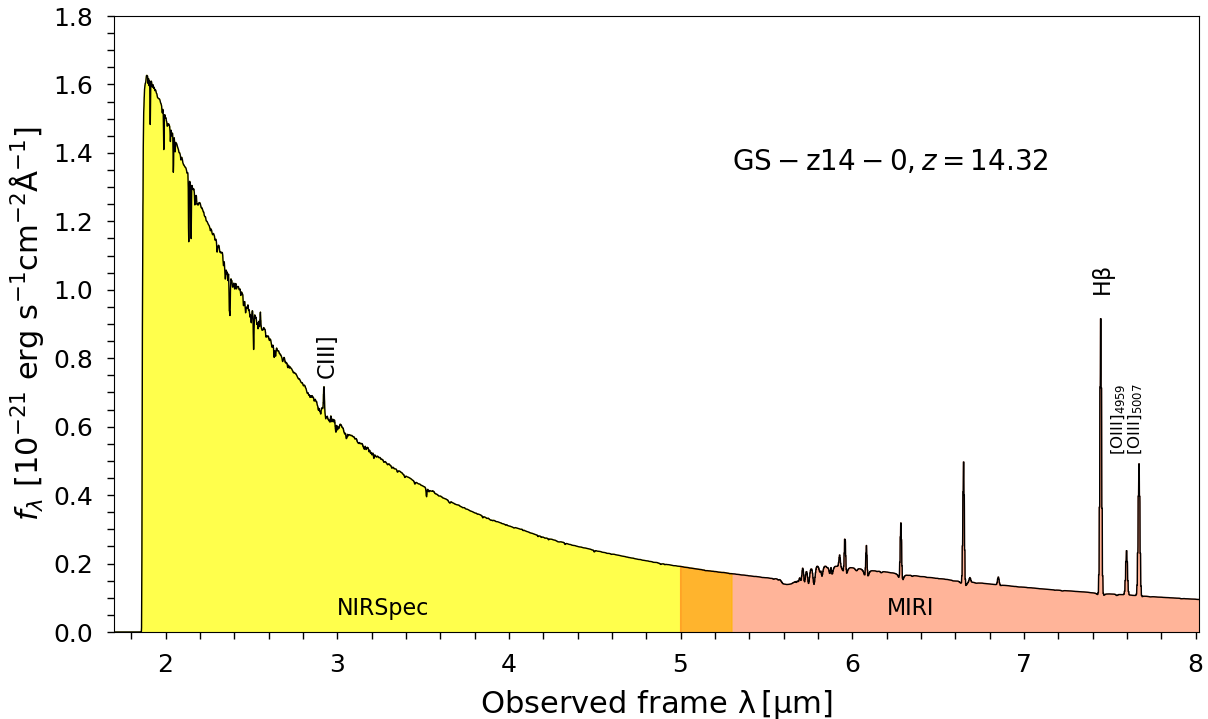}
\caption{Predicted spectrum of GS-z14-0 based on the SFH of the galaxy shown in Fig. \ref{Fig:SFH}, and for an ionization parameter $U=-1.2$. The spectrum has been computed using \code{BAGPIPES} \citep{Carnall18}. It has a UV spectral slope $\beta = -2.25$; the position of the $\rm CIII]\lambda1907, 1909$ line is indicated, along with those potentially detected by MIRI  \citep{Helton24}.} 
\label{Fig:Spectrum}
\end{figure*}

\paragraph{(e) The super-Eddington outflow phase}
Given that GS-z14-0 is observed at $z=14.32$ to be bright and blue (UV slope $\beta=-2.2$), the previous obscured phase must have come to an end. Indeed, the model predicts that during phase (d) the galaxy became super-Eddington at $\tl \approx 66$ Myr. Thus, after a time $t_D$, or about 26 Myr before observation, a powerful outflow develops inducing a dramatic decrease in the gas and dust content. As a consequence the SFR drops from $100$ to $15\, \msunyr$, and slightly rising thereafter. We pause to underline that, differently from LW feedback which was inducing a bursty behavior, radiation-driven outflows do not quench SFR completely, but they rather gradually decrease and control it. 

Outflows play also another fundamental role: by dispersing dust (and metals) \textit{they lower the dust opacity, thus making the galaxy bluer and bright.} In the absence of outflows, the dust mass ($\approx 10^6\, \msun$) synthesized by the observed stars in GS-z14-0 before the outflow, would be still co-located with stars, i.e. within $r_e \approx 145$ pc. Such high dust surface density of $ \approx 5\times 10^7\, \msun \rm kpc^{-2}$) would produce an $A_V \simeq 100$. Outflows instead push the dust (and gas) to large radii and into the circumgalactic medium, without necessarily expelling it beyond $r_{\rm vir}$, but dramatically decreasing its optical depth and $A_V$. As a result, they make the galaxy brighter -- naturally explaining \citep{Ferrara24} the overabundance of super-early galaxies deduced from the UV luminosity functions \citep[see, e.g.][and references therein]{McLeod23} -- and bluer, as observed. 

\paragraph{(f) The post-starburst, observed phase}
Following the reduction of the SFR, the galaxy settles into a sub-Eddington regime\footnote{In spite of the high observed sSFR$= 45^{+56}_{-35}\, \rm Gyr^{-1}$, due to its low $\cal D$, according to eq. \ref{eq:ssfr_thresh}, GS-z14-0 is \textit{not} super-Eddington.} and the outflow ceases. Contrary to the LW feedback recovery phase, the subsequent SFR increase is gradual. This is because the galaxy has been significantly devoided by its gas by the outflow, and it has to rely on accretion (acting on longer timescales) to replenish its reservoir. We conclude that GS-z14-0 is seen in such a post-starburst (or better, post-outflow) phase. 

Our model nicely reproduces all the observed quantities (see Tab. \ref{tab:properties}), predicting $\rm SFR =19.6\, \msunyr$, $\log (M_\star/\msun)=8.95$, $A_V = 0.34$, $\log (Z/\zsun)= -2.2$. We stress that this agreement should not be over-emphasized as at least some of the \quotes{observed} values have been deduced from a SED fitting procedure which requires specific assumptions about the shape of the SFH, which is in general different from the one predicted by our model.  

When normalized to final halo mass, the mean conversion efficiency of the cosmologically available gas ($f_b M$) into stars is $\langle \epsilon_\star \rangle = 0.7$\%. The mass-weighted age of the stellar population is $\langle t \rangle_M = 36$ Myr, i.e. only slightly larger than the free-fall time $t_{\rm ff}(z=14.32)$; such quantity contains important information on the non-monotonic SFH of the galaxy. The luminosity-weighted age is instead, as expected, shorter, $\langle t \rangle_L = 18.9$ Myr. We finally predict the $F_{4200}/F_{3500}$ Balmer break to be $B=2.02$.

\subsection{Predicted spectrum}
Using the SFH shown in Fig. \ref{Fig:SFH}, complemented with the predicted ionization parameter, $U\approx -1.2$,  we finally calculate the observed spectrum of GS-z14-0, shown in Fig. \ref{Fig:Spectrum}. To this aim we use the SED fitting code \code{BAGPIPES} \citep{Carnall18}. Due to the low metallicity, the spectrum is essentially featureless, showing no prominent nebular lines, with the only exception of a weak $\rm CIII]\lambda1907, 1909$ line with an equivalent width EW $\approx 4.4$ \AA. The UV spectral slope is $\beta = -2.25$, in excellent agreement with the observed one $\beta =-2.2\pm 0.07$. 

The spectrum in the MIRI band show luminous $\rm H\beta$ and $\rm [OIII]\lambda\lambda4959,5007$ lines, which are therefore likely responsible for the flux excess measured by \citet{Helton24} in the $7.7\ \mu$m filter. We note that, due to the low predicted  metallicity $Z\approx 0.01 Z_\odot$, the $\rm [OIII]/H\beta$ line ratio is also low $\approx 0.5$. Such ratio is nevertheless very sensitive to the precise value of $Z$. For example, if we artificially increase the predicted metallicity to $Z=0.03$, it increases to 1.6. 

\section{Discussion}
We have shown that the SFH of GS-z14-0, a luminous, blue galaxy at $z=14.32$, is complex and undergoes different phases. LW feedback, via \HH molecule photo-dissociation, induces a very bursty onset of star formation at $z \simeq 27$. As the galaxy grows the SFH becomes smoother, but the extremely compact size ($\approx 100$ pc) causes it to become heavily obscured and unobservable by JWST. It is only when the galaxy enters a super-Eddington phase that outflows can push the dust to large radii, reducing attenuation and allowing UV photons to be observed. This scenario has several implications.

\subsection{Stochasticity}
Our model confirms that there should be a considerable SFR stochasticity \citep{Mirocha23, Shen23, Pallottini23, Sun23, Ciesla23} in early galaxies due to LW feedback and outflows. However, higher fluctuations are expected during the initial phases when LW dominates and star formation is completely choked for tens of Myr. Outflows produce less extreme, by yet rapid, changes (see below) in the SFR once the galaxy becomes super-Eddington. 

\subsection{Obscured star formation}
We find that a substantial fraction of the stellar mass build-up process might be obscured. For GS-z14-0, we find that 61\% of its stars were formed with $A_V > 3$. This phase lasted for about 30-40 Myr, corresponding to $\approx 20\%$ of its active star forming phase, and was terminated by the onset of the outflow. 

These results imply that a substantial number of FIR sources should be present at high-$z$. Just before the beginning of the outflow phase ($z=15.3$), with $M_d \simeq 10^6 \, \msun$, a simple estimate for the expected continuum emission at restframe 88$\mu$m for GS-z14-0 gives a flux $F_{88} = (12, 31, 54)\ \mu$Jy for a dust temperature $T_d = (60, 80, 100)$ K, respectively. These flux levels can be detected by ALMA in a reasonable observation time\footnote{A potential challenge is nevertheless represented by the CMB, whose high temperature, $T_{\rm CMB}(z=14.32)= 41.8$ K, might partially or even entirely suppress the dust signal \citep{daCunha13, Sommovigo22}}.  Hints of such an obscured galaxy population might have already been found in the candidates discussed by \citet{Rodighiero23}. At the observed redshift ($z=14.32$) the dust \textit{in the galaxy} has decreased to a mere $M_d = 3\times 10^4\, \msun$. Even for $T_d =100$ K, $F_{88} < 2\ \mu$Jy. As most of the dust has been ejected in the circumgalactic medium, where $T_d$ might be close to the CMB temperature, its emission might be harder to detect. More detailed calculations are required to assess the detectability in this case. 

\subsection{Smouldering star formation}
Radiation-driven outflows are launched immediately after the galaxy luminosity crosses the Eddington limit. In our treatment this is equivalent to impose $\rm sSFR > sSFR^\star$. Although the precise value of the threshold is uncertain, it is reasonable to assume that $\rm sSFR^\star \propto 1/{\cal D}$ as the driving efficiency depends on the momentum transfer from UV photons to dust grains. Thus, as dust is carried out of the galaxy, the outflow fades away because  $\rm sSFR^\star$ increases and the SFR starves due to the decreasing availability of gas. 

This is seen in the sudden $7\times$ drop of the SFR in only 12 Myr, at $14 < t_{\rm look}/\rm Myr < 26$. Such \textit{fast suppression}  of the SFR is almost impossible to achieve with SN feedback \citep{Gelli24}. For this reason, radiation-driven outflows have been invoked to explain the properties of high-$z$ quiescent galaxies \citep{Gelli23, Carnall23, Looser23}.  

Interestingly, \citet[see their Fig. 20]{Robertson23} find that roughly half of the objects in their JADES Origin Field survey indicate a peak or burst in their star formation rates $\approx 10$ Myr before the observation epoch. This scenario is tantalizingly close to the one we are proposing here, with a an episode of “mini-quenching” following the onset of the radiation-driven outflow.

\subsection{Blue monsters}
Once the galaxy enters the post-starburst phase it is seen as a \quotes{blue monster}, namely a blue, luminous galaxy in which the light from young stars is basically unattenuated. The associated brightness increase (in spite of the fact that the post-starburst phase corresponds to a \textit{decrease} of the SFR with respect to the obscured phase)  is crucial to explain the overabundance of luminous, super-early galaxies \citep{Ferrara23a, Fiore23, Ziparo23, Ferrara24} 
deduced from the non-evolving bright-end of the UV LF \citep{McLeod23, Finkelstein23, Donnan24}. 

We note that in the post-starburst phase the metallicity also decreases. For GS-z14-0 we predict a metallicity $\simlt 0.01\, \zsun$, which might explain the absence of strong emission lines observed in this \citep{Carniani24} and other \citep{Curtis23} sources at $z>10$. The case of other galaxies like GN-z11 \citep{Bunker23} or GHZ2 \citep{Castellano24}, which show prominent UV/optical lines, might instead indicate that these systems are still at the beginning of the post-starburst phase or that the outflow has been less efficient.  

\subsection{Stellar ages}
GS-z14-0 forms 90\% of its stars in the last 50 Myr (Fig. \ref{Fig:SFH}), with a mass-weighted stellar age of 36 Myr, only slightly larger than the free-fall time at the observed redshift $z=14.32$. The SFH is in general increasing with time, following the growth of the dark matter halo, but there are periods in which the SFR is either quenched or smouldering. During these periods the SFR might be instead decreasing with time. This on-monotonic behavior might bias low age estimates using constantly increasing (or constant) priors on the SFH. However, non-parametric SED fitting procedures might well catch these more complex SFHs (for an example, see \citealt{Robertson23}). The presence of  moderately older stars (e.g., those formed in the obscured phase 25-70 Myr before observation) results in a relatively high Balmer break, $B\approx 2$. This indicator might be important to constrain the likely complex nature of SFH in super-early galaxies \citep{Vikaeus24,Borsani24,Langeroodi24}. At $z>10$ this measurement will have to rely on  \textit{JWST} MIRI observations. 




\section{Summary}
We have modelled the SFH of super-early galaxies, and applied the results to the most distant, spectroscopically confirmed galaxy GS-z14-0 at $z=14.32$. The main results are:

\begin{itemize}
\item[{\color{red} $\blacksquare$}] The SFH, starting at $z=26.7$, is complex. The initial ($z>18$), LW feedback-regulated phases are bursty, relatively faint (reaching $M_{\rm UV}=-18.4$) and unattenuated, hence at reach of deep \textit{JWST} surveys. When dust shielding enables a smooth SFR, the galaxy quickly gets heavily obscured. In such an obscured phase, lasting for $\approx 20\%$ of the entire active star forming time, about 70\% of all its stars are formed. Super-early galaxies in this phase should be detectable by ALMA. 
\item[{\color{red} $\blacksquare$}] As the galaxy becomes super-Eddington, 26 Myr prior to observation, a powerful radiation-driven outflow clears most of the dust levitating it into the circumgalactic medium, and reduces the metallicity to $Z\approx 0.01\ \zsun$. The outflow also \quotes{mini-quenches} the SFR by 7 times, from $100 \to 15\ \msunyr$. The galaxy becomes a \quotes{blue monster}, thus contributing to the enhancement of the bright-end of the UV LF.
\item[{\color{red} $\blacksquare$}] As the outflow ceases due to decreased dust opacity, the galaxy relaxes into a post-starburst phase in which it is currently observed \citep{Carniani24}. Our model nicely reproduces all the observed quantities (see Tab. \ref{tab:properties}), predicting $\rm SFR =19.6\, \msunyr$, $\log (M_\star/\msun)=8.95$, $A_V = 0.34$, $\log (Z/\zsun)= -2.2$. The mass-weighted age of the stellar population is $\langle t \rangle_M = 36$ Myr. We finally predict the $F_{4200}/F_{3500}$ Balmer break to be $B=2.02$.    
\end{itemize}


\acknowledgments
We thank S. Carniani, C. Conselice, S. Gallerani, M. Kohandel, V. Markov, P. Oesch, A. Pallottini, E. Parlanti for useful discussions, data and comments. This work is supported by the ERC Advanced Grant INTERSTELLAR H2020/740120. 
Plots in this paper produced with the \textsc{matplotlib} \citep{Hunter07} package for \textsc{PYTHON}.    

\bibliographystyle{aasjournal}
\bibliography{paper}

\end{document}